# Comment on "Influence of the Dzyaloshinskii-Moriya Exchange Interaction on Quantum Phase Interference of Spins"

E. del Barco[1,*], S. Hill[2] and D. N. Hendrickson[3]

[1] *Department of Physics, University of Central Florida, Orlando, FL 32816, USA*

[2] *National High Magnetic Field Laboratory and Department of Physics, Florida State University, Tallahassee, FL 32310, USA*

[3] *Department of Chemistry and Biochemistry, University of California at San Diego, La Jolla, CA 92093, USA*

[*] *Electronic mail: delbarco@physics.ucf.edu*

In a recent Letter [1], Wernsdorfer *et al*. report an experimental study of a $Mn_{12}$ molecular wheel which shows essentially identical behavior to the $Mn_{12}$ wheel studied by Ramsey *et al*. [2]. In their Letter, Wernsdorfer *et al*. use the same model of a dimer of two exchange-coupled spins used in [2] as a basis to extend the study of the influence of the Dzyaloshinskii-Moriya (DM) interaction on the quantum tunneling of the magnetization of this system; in particular, they show that a tilt of the DM vector away from the uniaxial anisotropy axis can account for the asymmetric nature of the quantum interference minima associated with resonances between states of opposite parity, e.g. $k = 1(A)$. We want to stress that the inclusion of DM interactions in a system with inversion symmetry cannot mix states of opposite parity, i.e. the parity operator commutes with the Hamiltonian. Consequently, the use by Wernsdorfer *et al.* of a single DM vector in a centrosymmetric dimer is strictly forbidden since it implicitly violates parity conservation. The authors correctly point out that the lack of an inversion center between each pair of manganese ions on the wheel justifies the possibility of local DM interactions, even though the complete molecule has an inversion center. However, these local DM interactions must also satisfy the molecular inversion symmetry, i.e. they cannot mix states of opposite parity. We agree that such DM interactions are not always completely innocuous, e.g. they can mix spin states having the *same* parity. Indeed, in Kagome systems [3] (cited in [1]), this can lead to weak ferromagnetism. Nevertheless, the inversion symmetry of the lattice is preserved and parity is still conserved.

DM interactions will not mix states of opposite parity unless the global inversion symmetry of the lattice is broken. Stochastic disorder could certainly be a source of local symmetry breaking, although this would result in a random distribution of DM vector orientations and magnitudes. Note that this is inconsistent with the observation of clear Berry phase oscillations, which would require an identical DM interaction and, therefore, an identical distortion of *all* molecules in the crystal. X-ray diffraction data indicate that the molecular inversion symmetry is preserved down to 100 K [4]. The possibility of a structural transition below 100 K cannot be ruled out, although magnetic and thermodynamic studies do not support such a hypothesis.

Finally, we would like to stress the fact that the $k = 1(A)$ tunnel splitting is substantially affected by the angle $\varphi_{DM}$ between the transverse projection of the tilted DM vector and the original hard anisotropy axis of the molecule which, according to the authors, "hardly depends on $\varphi_{DM}$". Indeed, for a tilt of the DM vector of $\theta_{DM} = 10^o$ (optimum value in [1]), the $k = 1(A)$ minima vanish almost completely for $\varphi_{DM} = 90^o$, when fixing the transverse field along $x$ (see Fig. 1a). This is because, for $\varphi_{DM} \neq 0$, the transverse projections of the DM vector and the transverse single-ion anisotropies are incommensurate, forcing different minima in $k = 1$(A) to occur at different transverse field orientations, as can be seen in Fig. 1b for the case of a moderate angle, $\varphi_{DM} = 30^o$. Indeed, this effect could be behind the rounding of the $k = 1$(A) Berry phase minima observed in Refs. [1,2]. One could re-orient the transverse field in order to recover sharp $k = 1$(A) minima. An experimental verification of this behavior would lead to a better understanding of the phenomena observed in this system.

**Figure and Figure Caption:**

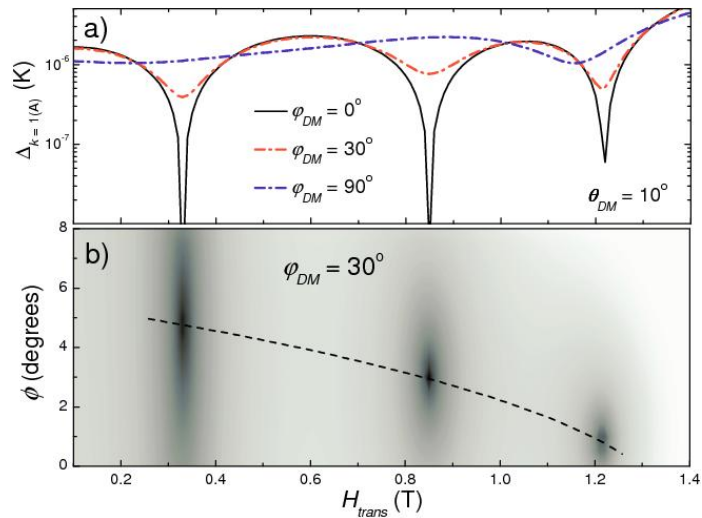

**Figure 1:** (Color online) a) Calculated tunnel splittings of resonance $k = 1(A)$ as a function of the transverse magnetic field, $H_x$, for different orientations of the DM vector; all other parameters are the same as those reported in [1]. b) Contour plot of the tunnel splitting as a function of the angle (*y*-axis) and magnitude (*x*-axis) of a transverse field for the azimuthal angle $\varphi_{DM} = 30°$. The dashed line connects the positions of the interference minima.